\documentclass[journal, 10pt]{IEEEtran}

\usepackage{multirow}
\usepackage[font=scriptsize,caption=false,labelsep=space]{subfig}
\usepackage{mathrsfs}
\usepackage{graphicx,float,wrapfig,epstopdf,amsmath}
\usepackage[]{algorithmicx}
\usepackage{algpseudocode,algorithm}
\usepackage{balance}
\usepackage{mathtools,lipsum}
\usepackage{amsmath}
\usepackage{amssymb}
\usepackage{amsthm}
\usepackage{graphicx}
\usepackage{epstopdf}
\usepackage{times}
\usepackage{textcomp,cite}

\usepackage{url}
\usepackage{hyperref}

\usepackage{multirow}
\usepackage{threeparttable}
\graphicspath{{./figures/}}

\DeclareMathOperator*{\argmax}{argmax} 
\DeclareMathOperator*{\maximize}{maximize} 
\DeclareMathOperator*{\subjectto}{subject \hspace{3pt} to:\hspace{3pt}} 

\usepackage[table]{xcolor}
\definecolor{intnull}{RGB}{213,229,255}
\definecolor{inteins}{RGB}{128,179,255}
\definecolor{color1}{RGB}{199,209,232}
\definecolor{color2}{RGB}{230,231,233}

\begin{document}
	\title{Hybrid Beamforming for Integrated Sensing and Communications With Low Resolution DACs   }
	
	\author{\IEEEauthorblockN{Ahmet M. Elbir, \textit{Senior Member, IEEE},  Abdulkadir Celik, \textit{Senior Member, IEEE}, \\ and Ahmed M. Eltawil, \textit{Senior Member, IEEE}  }

		\thanks{A. M. Elbir is with Dept. of Electrical and Electronics Engineering, Istinye University, Istanbul 34396, Turkey; and with King Abdullah University of Science and Technology, Thuwal 23955, Saudi Arabia			(e-mail: ahmetmelbir@ieee.org).}
		\thanks{A. {C}elik and A. M. Eltawil are with King Abdullah University of Science and Technology, Thuwal 23955, Saudi Arabia (e-mail:abdulkadir.celik@kaust.edu.sa, ahmed.eltawil@kaust.edu.sa). } 
		
	}
	\maketitle
	\begin{abstract}
		Integrated sensing and communications (ISAC) has emerged as a means to efficiently utilize spectrum and thereby save cost and power. At the higher end of the spectrum, ISAC systems operate at wideband using large  antenna arrays to meet the stringent demands for high-resolution sensing and enhanced communications capacity. On the other hand, the overall design should satisfy energy-efficiency and hardware constraints such as operating on low resolution components for a practical scenario. Therefore, this paper presents the design of \underline{H}ybrid \underline{AN}alog and \underline{D}igital \underline{B}e\underline{A}mformers with \underline{L}ow reso\underline{L}ution (HANDBALL) digital-to-analog converters (DACs). {We introduce a greedy-search-based approach to design the analog beamformers for multi-user multi-target ISAC scenario. Then, the quantization distortion is taken into account in order to design the baseband beamformer with low resolution DACs.} We evaluated performance of the proposed HANDBALL technique in terms of both spectral efficiency and sensing beampattern, providing a satisfactory sensing and communication performance for both one-bit and few-bit designs.	
	\end{abstract}


	\section{Introduction}
	\IEEEPARstart{I}{ntegrated} sensing and communications (ISAC) is considered as one of the vital technologies for deploying the next generation wireless networks as it provides joint sensing and communications (S\&C) functionalities in a common hardware over a shared spectrum~\cite{jrc_TCOM_Liu2020Feb}. This has led to a great amount of research efforts by both academia and industry on the implementations and standardization activities of ISAC. The ISAC platforms are expected to achieve satisfactory performance for both S\&C, which include achieving reliable communication rate while delivering sufficient power to illuminate the sensing targets~\cite{r3_Zhang2024Jan}.  Therefore, large antenna arrays are deployed at the transmitter to achieve high beamforming gain in massive multiple-input multiple-output (MIMO) configurations~\cite{elbir2022Nov_Beamforming_SPM}.
	
	At millimeter-wave (mmWave) frequencies, large antenna arrays allow signal transmission with high data rates thanks to large bandwidth and provide high resolution target resolution capabilities. Unfortunately, the high hardware cost and power consumption makes it difficult to realize fully digital (FD) data transmission with a dedicated radio-frequency (RF) chain per antenna in the array. To overcome this challenge, hybrid analog/digital beamformers are employed in order to achieve a cost-effective solution~\cite{lowRes_Mo2017Mar}. A common limitation of hybrid architectures is employing high resolution digital-to-analog converters (DACs), which consume high power at mmWave. For instance, a mmWave massive MIMO system with $512$ high resolution (8-12 bits) DACs, the total power consumption is about $256$ W~\cite{lowRes_mag_Zhang2018Apr}.  Thus, low resolution DACs (1-4 bits) lead to an alternative design to further reduce the power consumption as it grows exponentially with resolution~\cite{adc_survey_Walden1999Apr}.
	
	In the literature, there exist several works on the design of hybrid beamformers with low resolution data converters. For instance, low resolution DAC design is considered in~\cite{lowRes_Mo2017Mar} via both the inversion as well as the singular value decomposition (SVD) of the channel matrix for mmWave massive MIMO. Also in~\cite{lowRes_2_onebit_Hou2018Aug}, one-bit DACs are designed for hybrid beamforming based on Bussgang theorem. Similarly, an SVD-based approach is studied in~\cite{lowRes_3_EE_Ribeiro2018Apr} by investigating the  energy-efficiency (EE) of the overall MIMO system based on additive quantization noise model (AQNM). In~\cite{lowRes_oneBit_Maeng2020Aug}, an alternating optimization approach is proposed to design hybrid beamformers with both one-bit DAC and finite-quantized phase shifters.
	
	The aforementioned works mostly focus on communication-only systems~\cite{lowRes_Mo2017Mar,lowRes_2_onebit_Hou2018Aug,lowRes_3_EE_Ribeiro2018Apr,lowRes_oneBit_Maeng2020Aug}, whereas hybrid beamforming with low resolution data converters is rather unexamined for ISAC paradigm. A few recent works consider the ISAC scenario with FD processing rather than hybrid analog/digital beamforming. For instance, \cite{lowRes_EE_ISAC_RSMA_Dizdar2022Jun} and \cite{lowRes_ISAC_EE_LEO_Liu2024Feb} consider FD beamformers with low resolution DACs for rate-splitting multiple access (RSMA) in ISAC with EE optimization. {While hybrid beamforming is studied in~\cite{lowRes_HB_ISAC_Kaushik} for single-user multi-target ISAC, the quantized DAC output is only considered for communication-only beamformer design, which degrades the overall S\&C performance. }
	
	In this work,  we propose a  \underline{H}ybrid \underline{AN}alog and \underline{D}igital \underline{B}e\underline{A}mformers with \underline{L}ow reso\underline{L}ution (HANDBALL) approach for {multi-user multi-target ISAC scenario.} The proposed approach takes into account the covariance of the quantization distortion for both one-bit and few-bit designs based on both Bussgang Theorem~\cite{bussgang1952} and the AQNM~\cite{aqnm_Fletcher2007Dec,lowRes_Mo2017Mar}. {In order to design the analog beamformers, we introduce a greedy-search (GS)-based approach, wherein the overall analog beamformer is split into two portions: communication-only and sensing-only beamformers.  Then, we design the baseband beamformer with low resolution DAC by taking into account the quantization distortion.} Via numerical simulations, we show that our HANDBALL approach provides satisfactory performance in terms of both communication SE and the sensing beampattern.
	
	\textit{Notation:} Throughout the paper, {$(\cdot)^*$,} $(\cdot)^\textsf{T}$ and $(\cdot)^{\textsf{H}}$ denote the conjugate, transpose and conjugate transpose operations, respectively. For a matrix $\mathbf{A}$; $[\mathbf{A}]_{ij}$ and   {$[\mathbf{A}]_k$ correspond to the $(i,j)$-th entry and the $k$-th column} while $\mathbf{A}^{\dagger}$ denotes the Moore-Penrose pseudo-inverse of $\mathbf{A}$. A unit matrix of size $N$ is represented by $\mathbf{I}_N$, and  $\mathcal{Q} (\cdot)$  and $\mathrm{Tr}\{\cdot\}$ denote the quantization and trace operations, respectively. $|| \cdot ||_\mathcal{F}$ denotes the Frobenious norm while {$\odot$ and $\otimes$ denote Khatri-Rao and Kronecker products, respectively.} 
	\begin{figure}[t]
		\centering
		{\includegraphics[draft=false,width=\columnwidth]{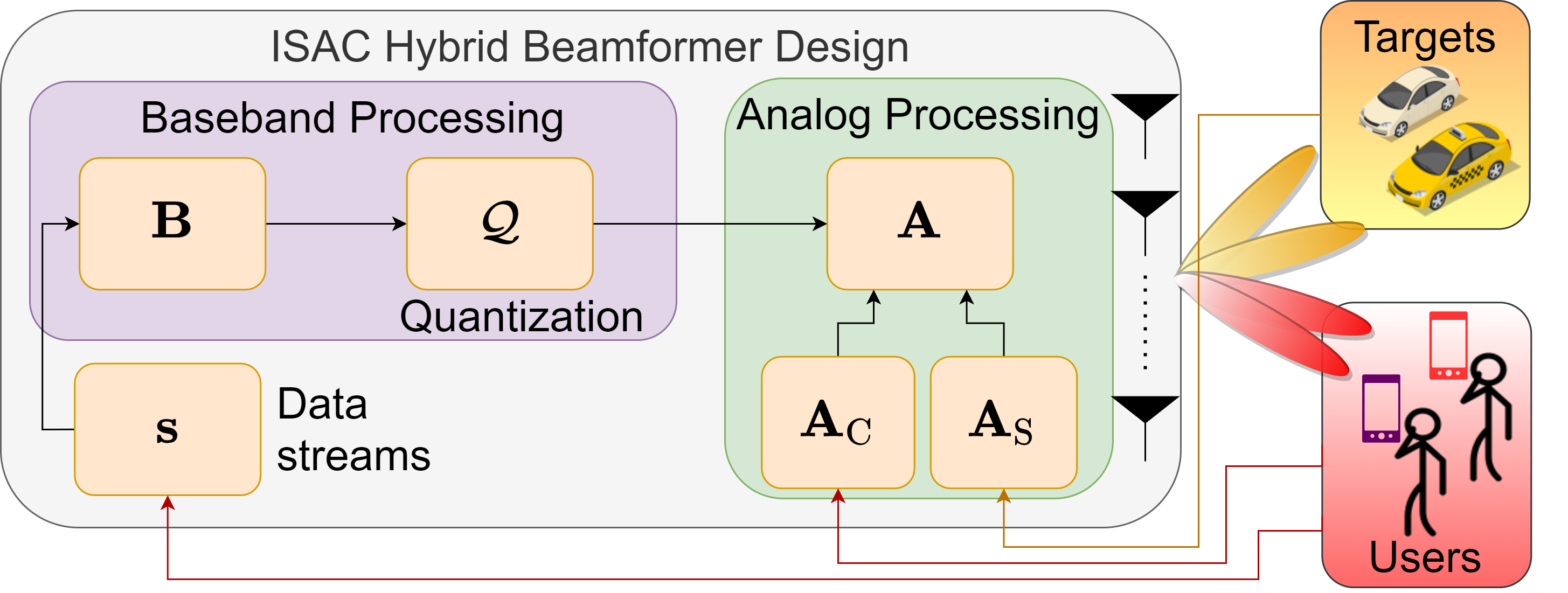} } 
		\caption{The ISAC hybrid beamforming architecture with  low-resolution DACs.
		}
		
		\label{fig_BS}
	\end{figure}
	
	\section{System Model \& Problem Formulation}
	Consider an ISAC system, as shown in Fig.~\ref{fig_BS}, wherein the dual-functional base station (DFBS) is equipped with $N_\mathrm{T}$ antennas and $N_\mathrm{RF}$ RF chains. { The DFBS simultaneously aims to generate multiple beams toward $T$ targets and communicate with $U$ users, which are equipped with $N_\mathrm{R}$ antennas, with $N_\mathrm{S}$ data-streams. To account for low power consumption and cheaper hardware, we assume that each user employs a single analog beamformer, hence a single data-stream is received per user, i.e, $N_\mathrm{S} = U$. In the downlink, the DFBS first applies a baseband beamformer $\mathbf{B} = [\mathbf{b}_{1},\cdots, \mathbf{b}_{U}]\in \mathbb{C}^{N_\mathrm{RF}\times U }$ to transmit the signal vector $\mathbf{s} = [s_1,\cdots, s_U]^\textsf{T}\in \mathbb{C}^{U}$, where $\mathbb{E}\{\mathbf{ss}^\textsf{H}\} = \frac{P_\mathrm{s}}{U}\mathbf{I}_U$. The precoded data is then quantized as $\mathcal{Q}(\mathbf{Bs})$. The DFBS applies analog beamformer $\mathbf{A}\in \mathbb{C}^{N_\mathrm{T}\times N_\mathrm{RF}}$, where the beamformer weights are realized with phase-shifters, i.e., the entries of $\mathbf{A}$ are subject to constant-modulus constraint as $|[\mathbf{A}]_{n_\mathrm{T},n_\mathrm{RF}}  | = 1/\sqrt{N_\mathrm{T}}$ for $ n_\mathrm{T} \in \{1,\cdots, N_\mathrm{T}\}$ and $ n_\mathrm{RF}\in \{1,\cdots, N_\mathrm{RF} \}$. Finally, the $N_\mathrm{T}\times 1$ transmit signal vector becomes ${\mathbf{x}} = \mathbf{A}\mathcal{Q}(\mathbf{Bs})  $, which travels through the mmWave channel $\mathbf{H}_u\in \mathbb{C}^{N_\mathrm{R}\times N_\mathrm{T}}$, which represents the superposition of $L$ non-line-of-sight (NLoS) paths for $u$-th user as		
		\begin{align}
			\mathbf{H}_u = \kappa \sum_{\ell = 1}^{L} \alpha_{u} \mathbf{a}_\mathrm{R}(\phi_{u,\ell}) \mathbf{a}_\mathrm{T}^\textsf{H}(\theta_{u,\ell}),
		\end{align}
		where $\kappa = \sqrt{\frac{N_\mathrm{R}N_\mathrm{T}}{L}}$ and $\alpha_u\in \mathbb{C}$ denotes complex channel gain. $\mathbf{a}_\mathrm{R}(\phi_{u,\ell})$ and $\mathbf{a}_\mathrm{T}(\theta_{u,\ell})$ represent the $N_\mathrm{R}\times 1$ and $N_\mathrm{T}\times 1$ steering vectors corresponding to the angle-of-arrival $\phi_{u,\ell}$ and angle-of-departure $\theta_{u,\ell}$ for the $\ell$-th scattering path of the $u$-th user, respectively.  	Assuming a uniform linear array (ULA) geometry with half-wavelength element spacing, the $n$-th element of $\mathbf{a}_\mathrm{T}(\theta_{u,\ell})$ is defined as 
			\begin{align}
			\left[\mathbf{a}_\mathrm{T}(\theta_{u,\ell})  \right]_n = \frac{1}{\sqrt{N_\mathrm{T}}}\exp\{- \mathrm{j} \pi (n-1) \sin\theta_{u,\ell}  \},
		\end{align}
	 and $\mathbf{a}_\mathrm{R}(\phi_{u,\ell})$ is defined similarly. At the user side, the $N_\mathrm{R}\times 1$  received signal at the $u$-th user becomes
			\begin{align}
			 \mathbf{y}_u = \mathbf{H}_u {\mathbf{x}} + \mathbf{n}_u,
		\end{align}
		 which is then  processed by the combiner $\mathbf{c}_u\in \mathbb{C}^{N_\mathrm{R}}$, where $|[\mathbf{c}]_{n_\mathrm{R}}  | = 1/\sqrt{N_\mathrm{R}}$ for $ n_\mathrm{R} \in \{1,\cdots, N_\mathrm{R}\}$. After combining, the received signal becomes
		\begin{align}
			\bar{y}_u = \mathbf{c}_u^\textsf{H}\left(\mathbf{H}_u {\mathbf{x}} + \mathbf{n}_u\right) = 	\mathbf{c}_u^\textsf{H}\mathbf{H}_u \mathbf{A}\mathcal{Q}(\mathbf{B}\mathbf{s})+ \mathbf{c}_u^\textsf{H}\mathbf{n}_u,
			\label{sigModel}
		\end{align}
		where  $\mathbf{n}_u\in\mathbb{C}^{N_\mathrm{R}}$ denotes the complex Gaussian noise with zero-mean and variance $\sigma_n^{2}$. 	 The DFBS also senses the $T$ targets in the environment. Define $\varphi_t$ as the LoS direction of the $t$-th target. Then, the corresponding steering vector of the $t$-th target is defined as $\mathbf{a}_\mathrm{T}(\varphi_t)\in \mathbb{C}^{N_\mathrm{T}}$.   Furthermore,  the $N_\mathrm{T}\times 1$ array output collected from the targets is given by 
		\begin{align}
			\tilde{\mathbf{y}} = \sum_{t = 1}^{T} \beta_t \mathbf{a}_\mathrm{T}(\varphi_t) \tilde{s}_{t} + \tilde{\mathbf{n}}, \label{sigSensing}
		\end{align}
		where $\beta_t\in \mathbb{C}$ and $\tilde{s}_{t}\in \mathbb{C}$ denote the radar cross-section and the echo signal for the $t$-target, respectively, and  $\tilde{\mathbf{n}}\sim \mathcal{N}(0,\sigma_n^2)$ represents the Gaussian noise vector.

		The aim of this work is to design the ISAC hybrid beamformers $\mathbf{A}$, $\mathbf{B}$ and $\mathbf{C} = [\mathbf{c}_1,\cdots, \mathbf{c}_U]\in\mathbb{C}^{N_\mathrm{R}\times U}$ to simultaneously communicate with $U$ users and sense $T$ targets while taking into account the quantized baseband signals. To this end, the hybrid beamformers are optimized via maximizing the sum-rate of the overall system $\mathcal{R}$, which is given by 
		\begin{align}
			\mathcal{R} = \sum_{u = 1}^{U} \log_2 (1 + \mathrm{SIQNR}_u),
		\end{align}
		where $ \mathrm{SIQNR}_u$ is defined as the signal to interference, quantization and noise ratio (SIQNR), which is defined as
		\begin{align}
			\mathrm{SIQNR}_u =  \frac{\frac{P_s}{U}  |\mathbf{c}_u^\textsf{H}\mathbf{H}_u \mathbf{A} \mathcal{Q}(\mathbf{b}_u s_u)      |^2        }{ \frac{P_s}{U}\sum_{k \neq u}|\mathbf{c}_u^\textsf{H}\mathbf{H}_u \mathbf{A} \mathcal{Q}(\mathbf{b}_k s_k)      |^2 + \sigma_n^2  }.
		\end{align}
		Then, we can write the  hybrid beamformer design problem as
		\begin{subequations}
			\begin{align}
				&\maximize_{\mathbf{A},\mathbf{B},\mathbf{C}} \; \mathcal{R} \nonumber \\
				& \subjectto    \|  \mathbf{A} \mathcal{Q} (\mathbf{Bs}  )  \|_\mathcal{F}^2 \leq P_\mathrm{max}, \label{c1} \\
				& \hspace{50pt}  |[\mathbf{A}]_{n_\mathrm{T},n_\mathrm{RF}}  | = 1/\sqrt{N_\mathrm{T}}, \forall n_\mathrm{T},n_\mathrm{RF}, \label{c2} \\ 
				& \hspace{50pt}  |[\mathbf{C}]_{n_\mathrm{R},u}  | = 1/\sqrt{N_\mathrm{R}}, \forall n_\mathrm{R},u, \label{c3}
			\end{align}
			\label{problem1}
		\end{subequations}
		where $P_\mathrm{max}$ denotes the maximum transmit power. In (\ref{problem1}), in the constraints in (\ref{c1}), (\ref{c2}) and (\ref{c3}) are due to maximum transmit power $P_\mathrm{max}$ and the constant-modulus structure of analog beamformers $\mathbf{A}$ and $\mathbf{C}$, respectively. The optimization problem in (\ref{problem1}) is non-convex and non-linear due to multiple unknowns $\mathbf{A}$, $\mathbf{B}$ and $\mathbf{C}$, and quantization operation $\mathcal{Q}(\mathbf{b}_u s_u)$. In order to effectively solve (\ref{problem1}), we present a GS-based approach in the following.

	}	
	
	{
		\section{Proposed Method: HANDBALL}
		Our HANDBALL approach comprises two parts: 1) Analog beamforming, wherein the communication-only and sensing-only beamformers are designed via a greedy-search given the channel matrix of the users and the array output collected from the target echoes; 2) Baseband beamforming, wherein the quantization model is taken into account for low-resolution DAC design.

		\subsection{Analog Beamformer Design}
		In order to provide an effective solution, the optimization problem in (\ref{problem1}) can be equivalently solved via sparse signal recovery techniques, wherein the aim is to search for the best representation of the beamformer components from the feasible set of beamformer candidates~\cite{mimoRHeath,heathLimitedFeedBackMultiUser}. Thus, we define $\mathcal{F}$ and $\mathcal{W}$ as the feasible sets of precoder and combiners, for which we have $\mathbf{A}\in \mathcal{F}$ and $\mathbf{C}\in \mathcal{W}$, respectively. Then, we rewrite the problem in  (\ref{problem1}) as 
		\begin{align}
			&\maximize_{\mathbf{A},\mathbf{B},\mathbf{C}} \; \mathcal{R} \nonumber \\
			& \subjectto   \|  \mathbf{A} \mathcal{Q} (\mathbf{Bs}  )  \|_\mathcal{F}^2 \leq P_\mathrm{max}, \nonumber \\
			& \hspace{50pt}  \mathbf{A} \in \mathcal{F}, \mathbf{C} \in \mathcal{W}.		\label{problem2}
		\end{align}
		In order to maximize the sum-rate in (\ref{problem2}), we present a GS-based approach, wherein the signal power of each user is maximized based on the feasible sets $\mathcal{F}$ and $\mathcal{W}$. Furthermore, we design the analog beamformer $\mathbf{A}$ with two portions as
		\begin{align}
			\mathbf{A} = \left[\mathbf{A}_\mathrm{C}\; \mathbf{A}_\mathrm{S}\right],
		\end{align}
		where 
		$\mathbf{A}_\mathrm{C}\in \mathbb{C}^{N_\mathrm{T}\times U}$ and $\mathbf{A}_\mathrm{S}\in \mathbb{C}^{N_\mathrm{T}\times T}$ as the communication-only and sensing-only analog beamformers, respectively. 	In the following, we present our GS-based approach to design beamformers for communication ($\mathbf{A}_\mathrm{C}$ and $\mathbf{C}$) and sensing ($\mathbf{A}_\mathrm{S}$).

		\subsubsection{Design for $\mathbf{A}_\mathrm{C}$ and $\mathbf{C}$} Define $\mathbf{\Psi}_\mathbf{A}\in \mathbb{C}^{N_\mathrm{T}\times K}$ and $\mathbf{\Psi}_\mathbf{C}\in \mathbb{C}^{N_\mathrm{R}\times K}$ as the dictionary matrices from the feasible sets $\mathcal{F}$ and $\mathcal{W}$, respectively. In particular, $\mathbf{\Psi}_\mathbf{A}$ and $\mathbf{\Psi}_\mathbf{C}$ are composed of steering vectors $\mathbf{a}_\mathrm{T}(\theta^k)\in \mathbb{C}^{N_\mathrm{T}}$ and $\mathbf{a}_\mathrm{R}(\phi^k)\in \mathbb{C}^{N_\mathrm{R}}$ as $\mathbf{\Psi}_\mathbf{A} = \left[\mathbf{a}_\mathrm{T}(\theta^1),\cdots, \mathbf{a}_\mathrm{T}(\theta^K) \right]$ and  $\mathbf{\Psi}_\mathbf{C} = \left[ \mathbf{a}_\mathrm{R}(\phi^1),\cdots, \mathbf{a}_\mathrm{R}(\phi^K) \right]$, for the direction set $\theta^k, \phi^k \in \Theta = [0,\frac{2\pi}{K},\frac{4\pi}{K},\cdots, \frac{(K-1)2\pi}{K}]$, respectively. Next, we define the complete dictionary $\mathbf{\Psi}\in \mathbb{C}^{N_\mathrm{R}N_\mathrm{T}\times K^2}$ as 
		\begin{align}
			\mathbf{\Psi} = \mathbf{\Psi}_\mathbf{A}^*\odot\mathbf{\Psi}_\mathbf{C}.
		\end{align}
		In order to maximize the signal power for each user, we aim to maximize the correlation between the analog beamformer candidate and the unconstrained beamformers. Thus, we first define $\mathbf{g}_u = \mathbf{f}_{\mathrm{opt},u}^*\otimes \mathbf{v}_{\mathrm{opt},u}$, as the overall unconstrained beamformer, where $\mathbf{f}_{\mathrm{opt},u}\in \mathbb{C}^{N_\mathrm{T}}$ and $\mathbf{v}_{\mathrm{opt},u}\in \mathbb{C}^{N_\mathrm{R}}$ represent the unconstrained precoder/combiners for the $u$-th user, respectively. In particular, $\mathbf{f}_{\mathrm{opt},u}$ can be obtained from the singular value decomposition (SVD) of the channel  $\mathbf{H}_u$~\cite{elbir2022Nov_Beamforming_SPM}. Similarly, $\mathbf{v}_{\mathrm{opt},u}$ is defined as 
		\begin{align}
			\mathbf{v}_{\mathrm{opt},u} = \frac{1}{P_s} \bigg(\mathbf{f}_{\mathrm{opt},u}^\textsf{H} \mathbf{H}_u^\textsf{H} \mathbf{H}_u \mathbf{f}_{\mathrm{opt},u} + \frac{ \sigma_n^2}{P_s} \bigg)^{-1} \mathbf{f}_{\mathrm{opt},u}^\textsf{H}\mathbf{H}_u. 
		\end{align}
		Then,  the $u$-th column of the analog beamformer pairs of $\mathbf{A}_\mathrm{C}$ and $\mathbf{C}$  can be found from the columns of dictionaries as $[\mathbf{\Psi}_\mathbf{A}]_{p^\star}$ and $[\mathbf{\Psi}_\mathbf{C}]_{q^\star}$ for the $u$-th user via
		\begin{align}
			\{p^\star, q^\star  \} = \argmax_{p,q} |\boldsymbol{\psi}_{p,q}^\textsf{H} \mathbf{g}_{u}  |,
		\end{align}
		where $\boldsymbol{\psi}_{p,q}\in \mathbb{C}^{N_\mathrm{R}N_\mathrm{T}}$ is defined as $\boldsymbol{\psi}_{p,q} = [\mathbf{\Psi}_\mathbf{A}]_p^* \otimes [\mathbf{\Psi}_\mathbf{C}]_q$.

		\subsubsection{Design for $\mathbf{A}_\mathrm{S}$}In order to design the sensing-only beamformers, the array output  of the DFBS in (\ref{sigSensing}), $\tilde{\mathbf{y}}$, is utilized. Then, the $t$-th column of the sensing-only analog beamformer $\mathbf{A}_\mathrm{S}$ is obtained as $[\mathbf{\Psi}_\mathbf{A}]_{z^\star} $ via 
		\begin{align}
			\{z^\star\}  = \argmax_{z} | \tilde{\mathbf{y}}^\textsf{H} [\mathbf{\Psi}_\mathbf{A}]_z |,
		\end{align}
		where  $[\mathbf{\Psi}_\mathbf{A}]_z $ denotes $z$-th column of the dictionary $\mathbf{\Psi}_\mathbf{A}$ for $z = 1,\cdots, K$.
		
		In Algorithm~\ref{alg_analog}, we present the algorithmic steps for the design of communication-only ($\mathbf{A}_\mathrm{C}$) and sensing-only ($\mathbf{A}_\mathrm{S}$) beamformers, respectively. The complexity of Algorithm~\ref{alg_analog} is mainly due to the computation of steps 3 ($O(N_\mathrm{T}N_\mathrm{R}[ N_\mathrm{T} + N_\mathrm{R}])$), 4 ($O(K^2N_\mathrm{R}N_\mathrm{T}  )$), 8 ($O(KN_\mathrm{T})$) and 10 ($O(N_\mathrm{T}^2[N_\mathrm{R} + 1])$).

		\begin{algorithm}[t]
			\begin{algorithmic}[1] 
				\caption{ \bf HANDBALL - Analog Beamforming }
				\Statex {\textbf{Input:}  \label{alg_analog} $\{\mathbf{H}_u, \mathbf{f}_{\mathrm{opt},u},\mathbf{v}_{\mathrm{opt},u}, \mathbf{y}_u\}_{u=1}^U$, $\tilde{\mathbf{y}}$, $\mathbf{\Psi}_\mathbf{A}$, $\mathbf{\Psi}_\mathbf{C}$, $\sigma_n^2$.}
				\State $\mathbf{A}_\mathrm{C} = \text{Empty}$, $\mathbf{A}_\mathrm{S} = \text{Empty}$, $\mathbf{C} = \text{Empty}$,  $\tilde{\mathbf{y}}_\mathrm{res} = \tilde{\mathbf{y}}$.
				\State \textbf{for} $u=1,\cdots, U$ \textbf{do}
				\State \indent $\mathbf{g}_u = \mathbf{f}_{\mathrm{opt},u}^* \otimes \mathbf{v}_{\mathrm{opt},u}  $,  $\boldsymbol{\psi}_{p,q} = [\mathbf{\Psi}_\mathbf{A}]_p^* \otimes [\mathbf{\Psi}_\mathbf{C}]_q$.
				\State \indent $\{p^\star, q^\star  \} = \argmax_{p,q} |\boldsymbol{\psi}_{p,q}^\textsf{H} \mathbf{g}_{u}  |.$
				\State \indent $\left[\mathbf{A}_\mathrm{C}\right]_u\gets \left[\mathbf{\Psi}_\mathrm{A}\right]_{p^\star}$, $\left[\mathbf{C}\right]_u \gets  \left[\mathbf{\Psi}_\mathrm{C}\right]_{q^\star}$.
				\State \textbf{end}
				\State \textbf{for} $t=1,\cdots, T$ \textbf{do}
				\State \indent $		\{z^\star  \} = \argmax_{z} | \tilde{\mathbf{y}}_\mathrm{res}^\textsf{H} [\mathbf{\Psi}_\mathbf{A}]_z | .$
				\State \indent $\left[\mathbf{A}_\mathrm{S}\right]_t\gets  \left[\mathbf{\Psi}_\mathrm{A}\right]_{z^\star}$.
				\State \indent  $\tilde{\mathbf{y}}_\mathrm{res} \gets  \tilde{\mathbf{y}}_\mathrm{res} - \left[\mathbf{\Psi}_\mathrm{A}\right]_{z^\star} (\left[\mathbf{\Psi}_\mathrm{A}\right]_{z^\star})^\dagger  \tilde{\mathbf{y}}_\mathrm{res}$.
				\State \textbf{end}
				\Statex \textbf{Return:} $\mathbf{A} = \left[\mathbf{A}_\mathrm{C}\; \mathbf{A}_\mathrm{S}\right]$.

			\end{algorithmic} 
		\end{algorithm}

	}
	
	{
		\subsection{Baseband Beamformer Design}
		In this part, we first introduce the quantization model for both few-bit and one-bit case, then design the beamformers.
	}

	\subsubsection{Quantization Model}
	
	Consider the received  signal in (\ref{sigModel}), where the quantization term $\mathcal{Q} (\mathbf{Bs})$ is defined via the AQNM~\cite{aqnm_Fletcher2007Dec,lowRes_Mo2017Mar} as
	\begin{align}
		\mathcal{Q}_\mathrm{A} (\mathbf{Bs}) \approx \mathbf{W}_\mathrm{A} \mathbf{Bs} + \mathbf{d}_{\mathrm{A}},
		\label{model_AQNM}
	\end{align}
	where $\mathbf{W}_\mathrm{A} = \sqrt{1 - \epsilon_b} \mathbf{I}_{N_\mathrm{RF}}$ is the $N_\mathrm{RF}\times N_\mathrm{RF}$ weighting matrix and $\mathbf{d}_{\mathrm{A}}\in \mathbb{C}^{N_\mathrm{RF}}$ denotes the quantization distortion with the $N_\mathrm{RF}\times N_\mathrm{RF}$ covariance
	\begin{align}
		\mathbf{\Sigma}_{\mathrm{A}} = \mathbb{E} \{\mathbf{d}_{\mathrm{A}} \mathbf{d}_{\mathrm{A}}^\textsf{H} \} = \frac{P_s \epsilon_b}{U  } \mathrm{diag}\{\mathbf{B}\mathbf{B}^\textsf{H}  \}, 
		\label{ca}
	\end{align}
	where $\epsilon_b$ represents the quantization distortion factor, and it can be approximated as $\epsilon_b \approx \frac{\pi \sqrt{3}}{2} 2^{-2b} $ for $b$-bit quantization~\cite{lowRes_Mo2017Mar}. Another quantization model for $b =1$ is based on the Bussgang Theorem~\cite{bussgang1952} that involves
	\begin{align}
		\mathcal{Q}_\mathrm{B} (\mathbf{B} \mathbf{s}) \approx \mathbf{W}_\mathrm{B} \mathbf{B}\mathbf{s} + \mathbf{d}_\mathrm{B},
		\label{model_Bussgang}
	\end{align}
	for which the corresponding weighting matrix  $\mathbf{W}_\mathrm{B}\in \mathbb{C}^{N_\mathrm{RF}\times N_\mathrm{RF}}$ is given by
	\begin{align}
		\mathbf{W}_\mathrm{B} \hspace{-2pt}= \hspace{-2pt}\sqrt{\hspace{-2pt}\frac{2}{\pi}} \hspace{-2pt}\left[ \mathrm{diag}\{\hspace{-1pt}\mathbf{\Sigma}_{{\mathbf{x}}}\hspace{-1pt}\} \right]^{-\frac{1}{2}}\hspace{-2pt} =\hspace{-2pt} \sqrt{\frac{2 U}{\pi P_s} }\hspace{-2pt} \left[ \mathrm{diag}\{\mathbf{B}\hspace{-1pt} \mathbf{B} \}   \right]^{-\frac{1}{2}}  ,
		\label{wb}
	\end{align}
	where $\mathbf{\Sigma}_{{\mathbf{x}}} = \mathbb{E} \{ \tilde{\mathbf{x}} \tilde{\mathbf{x}}^\textsf{H} \} \in \mathbb{C}^{N_\mathrm{RF}\times N_\mathrm{RF}}$ is the covariance of $\tilde{\mathbf{x}} = \mathbf{B}\mathbf{s}\in \mathbb{C}^{N_\mathrm{RF}}$, i.e., $\mathbf{\Sigma}_{{\mathbf{x}}} = \frac{P_s}{U}\mathbf{B}  \mathbf{B}^\textsf{H}\in \mathbb{C}^{N_\mathrm{RF}\times N_\mathrm{RF}}$. Furthermore, the quantization distortion vector $\mathbf{d}_\mathrm{B}\in \mathbb{C}^{N_\mathrm{RF}}$ has the following covariance
	\begin{align}
		\mathbf{\Sigma}_\mathrm{B} = \bar{\mathbf{\Sigma}}_\mathrm{B} - \mathbf{W}_\mathrm{B} \mathbf{\Sigma}_{{\mathbf{x}}} \mathbf{W}_\mathrm{B}, 
		\label{cb}
	\end{align}
	where $\bar{\mathbf{\Sigma}}_\mathrm{B}\in \mathbb{C}^{N_\mathrm{RF}\times N_\mathrm{RF}}$ is defined as
	\begin{align}
		&\mathbf{\bar{\mathbf{\Sigma}}}_\mathrm{B} = \frac{2}{\pi} \bigg[ \arcsin \mathrm{diag}\{\mathbf{\Sigma}_\mathbf{x} \}^{-1/2} \mathrm{Re}\{\mathbf{\Sigma}_{{\mathbf{x}}} \} \mathrm{diag}\{\mathbf{\Sigma}_\mathbf{x} \}^{-1/2}    \big) \nonumber \\ & + \mathrm{j}\arcsin \big(\mathrm{diag}\{\mathbf{\Sigma}_\mathbf{x} \}^{-1/2} \mathrm{Im}\{\mathbf{\Sigma}_{{\mathbf{x}}} \} \mathrm{diag}\{\mathbf{\Sigma}_\mathbf{x} \}^{-1/2}    \big) \bigg].
	\end{align}

	In comparison, the AQNM model in (\ref{model_AQNM}) ignores the correlation among the entries of distortion vector $\mathbf{d}_\mathrm{A}$ whereas the model in (\ref{model_Bussgang}) which is based on the Bussgang theorem accounts for the correlation of the entries of $\mathbf{d}_\mathrm{B}$, thereby achieving superior performance~\cite{aqnm_Fletcher2007Dec}. Furthermore, the AQNM can offer analytical tractability for few-bit quantization whereas the Bussgang theorem only characterizes one-bit quantization~\cite{lowRes_comparison_AWNM_BT_Zhang}. In the rest of this work, we drop the subscript $\mathrm{A}$, $\mathrm{B}$, and use $\mathcal{Q}$, $\mathbf{W}$ and $\mathbf{\Sigma} $ whenever they are applicable for beamformers, i.e., $\mathrm{B}$ for $b=1$ and $\mathrm{A}$ for $b>1$, respectively.

	{
		\subsubsection{Beamformer Design}
		Once the analog beamformer $\mathbf{A}$ is designed as in Algorithm~\ref{alg_analog}, we first compute effective channel  $\mathbf{H}_\mathrm{eff}\in \mathbb{C}^{U\times N_\mathrm{RF}}$ manage  the interference among the users, i.e., 
		\begin{align}
			\mathbf{H}_\mathrm{eff} = \left[
			(\mathbf{c}_1^\textsf{H} \mathbf{H}_1 \mathbf{A})^\textsf{T},\cdots,  (\mathbf{c}_U^\textsf{H} \mathbf{H}_U \mathbf{A})^\textsf{T}  \right]^\textsf{T}, \label{heff}
		\end{align}
		for which we can obtain the initial baseband beamformer is  $	\overline{\mathbf{B}} = \mathbf{H}_\mathrm{eff}^\dagger$.} {Next, we normalize the rows of $\overline{\mathbf{B}}$ to account for communication-sensing trade-off. Thus, we write $\overline{\mathbf{B}}$ as  $\overline{\mathbf{B}} = \left[\overline{\mathbf{B}}_\mathrm{C}^\textsf{T},\; \overline{\mathbf{B}}_\mathrm{S}^\textsf{T} \right]^\textsf{T}$, where $\overline{\mathbf{B}}_\mathrm{C}\in \mathbb{C}^{U\times U}$ and $\overline{\mathbf{B}}_\mathrm{S}\in \mathbb{C}^{T\times U}$ correspond to communication and sensing analog beamformers, respectively. Define $0\geq \eta \geq 1$ as the communication-sensing trade-off parameter, for which $\eta = 0$ and $\eta = 1$ correspond to sensing-only and communication-only design, respectively. Then, $\overline{\mathbf{B}}$ is normalized as
		\begin{align}
			\hat{\mathbf{B}} = \left[ \begin{array}{cc}
				\frac{\eta }{\|\overline{\mathbf{B}}_\mathrm{C} \|_\mathcal{F}  } \mathbf{I}_{U} & \mathbf{0} \\
				\mathbf{0} & \frac{1-\eta }{\|\overline{\mathbf{B}}_\mathrm{S} \|_\mathcal{F}    } \mathbf{I}_{T}
			\end{array} \right] \overline{\mathbf{B}}.
		\end{align}
	}

	Note that the final baseband beamformer should satisfy the power constraint in (\ref{c1}), which can be written as  
	\begin{align}
		\|  \mathbf{A}\mathcal{Q} (\mathbf{Bs})  \|_\mathcal{F}^2 &\leq P_\mathrm{max} \nonumber \\
		\frac{P_s}{U}\|\mathbf{A} \mathbf{W}\mathbf{B}  \|_\mathcal{F}^2 + \mathrm{Tr}\{\mathbf{A}\mathbf{\Sigma}\mathbf{A}^\textsf{H}  \}     & \leq  P_\mathrm{max} \nonumber \\
		\frac{P_s}{U}\|\mathbf{A} \mathbf{W}\mathbf{B}   \|_\mathcal{F}^2 + \mathrm{Tr}\{\mathbf{\Sigma}  \}& \leq  P_\mathrm{max}, \label{powerConstraint}
	\end{align}
	where we assume the semi-unitary property of $\mathbf{A}$, i.e., $\mathbf{A}^\textsf{H}\mathbf{A} = \mathbf{I}_{N_\mathrm{RF}}$, and {$\mathrm{Tr}\{\mathbf{A}\mathbf{\Sigma}{\mathbf{A}^\textsf{H}} \} = \mathrm{Tr}\{\mathbf{\mathbf{A}^\textsf{H}\mathbf{A}\mathbf{\Sigma}} \}$}~\cite{heathLimitedFeedBackMultiUser,lowRes_oneBit_Maeng2020Aug}. We therefore normalize the power of the baseband beamformer $\hat{\mathbf{B}}$  as  
	\begin{align}
		\mathbf{B} =  \left( \frac{{ P_\mathrm{max}} - \mathrm{Tr}\{\mathbf{\Sigma}  \} }{\frac{P_s}{U} \|\mathbf{W}\mathbf{B}   \|_\mathcal{F}^2   } \right)^{\frac{1}{2}} \hat{\mathbf{B}}.
	\end{align}

	\begin{algorithm}[t]
		\begin{algorithmic}[1] 
			\caption{ \bf HANDBALL - Baseband Beamforming }
			\Statex {\textbf{Input:} $\mathbf{A}$, $\mathbf{C}$, $\{\mathbf{H}_u\}_{u=1}^U$,  $b$,  $P_s$, $P_\mathrm{max}$.} \label{alg_bb}
			\State Compute $\mathbf{H}_\mathrm{eff}$ as in (\ref{heff}).
			\State $	\hat{\mathbf{B}} = \mathbf{H}_\mathrm{eff}^\dagger$.
			\State  \textbf{if} $b =1$ \textbf{do}
			\State \indent  Compute $\mathbf{W}_\mathrm{B} $ and $\mathbf{\Sigma}_\mathrm{B}$ from (\ref{wb}) and  (\ref{cb}). 
			\State \indent Update $\mathbf{B}$ as  $ \mathbf{B} =  \left( \frac{{ P_\mathrm{max}} - \mathrm{Tr}\{\mathbf{\Sigma}_\mathrm{B}  \} }{\frac{P_s}{U} \|\mathbf{W}_\mathrm{B}\mathbf{B}   \|_\mathcal{F}^2   } \right)^{\frac{1}{2}} \hat{\mathbf{B}} $.
			\State  \textbf{else if} $b >1$ \textbf{do}
			\State \indent  Compute $\mathbf{W}_\mathrm{A} $ and $\mathbf{\Sigma}_\mathrm{A}$ from (\ref{model_AQNM}) and (\ref{ca}).
			\State \indent Update $\mathbf{B}$ as  $ \mathbf{B} =  \left( \frac{{ P_\mathrm{max}} - \mathrm{Tr}\{\mathbf{\Sigma}_\mathrm{A}  \} }{\frac{P_s}{U} \|\mathbf{W}_\mathrm{A}\mathbf{B}   \|_\mathcal{F}^2   } \right)^{\frac{1}{2}}\hat{\mathbf{B}} $.
			\State  \textbf{end}
			\Statex \textbf{Return:} $\mathbf{B}$.
			
		\end{algorithmic} 
	\end{algorithm}

	In Algorithm~\ref{alg_bb}, we present the algorithmic steps of our HANDBALL technique for baseband beamforming for both one-bit (steps 3-5) and few-bit (steps 6-9) scenarios. The complexity of Algorithm~\ref{alg_bb} are mainly due to the computation of $\hat{\mathbf{B}}$ and $\mathbf{B}$ with the order of $O(2N_\mathrm{RF}^2U  )$ and $O(N_\mathrm{RF} (N_\mathrm{RF} + U))$, respectively.


	\section{Numerical Experiments}
	The performance of our HANDBALL approach is evaluated in comparison with FD beamformer with no inter-user interference in terms of SE as well as array beampattern. 
	During the numerical simulations, we select $N_\mathrm{T} = 128$, $N_\mathrm{R} = 10$, and $N_\mathrm{RF} = 8$.  We assume $U = 3$, $T = 3$ and $L=5$, where the directions of target and users are uniform randomly selected, i.e., $\phi_{u,\ell}, \theta_{u,\ell}, \varphi_t \sim \mathrm{unif}\{[0^\circ,180^\circ]\}$. Furthermore, we select $P_\mathrm{max} = P_\mathrm{s} = 1$.

	\begin{figure}[t]
		\centering
		{\includegraphics[draft=false,width=\columnwidth]{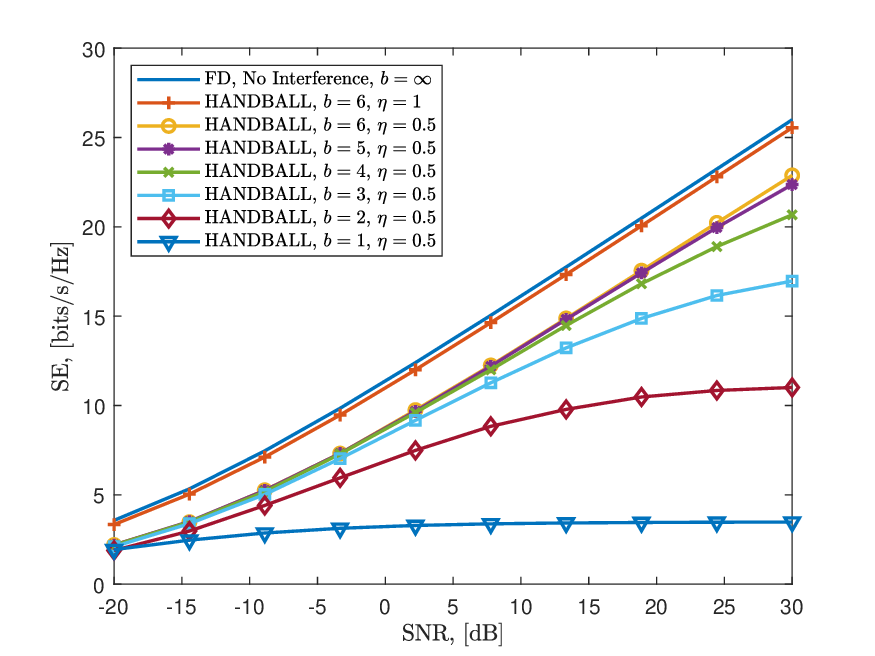} } 
		\caption{SE performance versus SNR when $U = 3$, $T = 3$.
		}
		
		\label{fig_SNR}
	\end{figure}

\begin{figure}[t]
		\centering
		{\includegraphics[draft=false,width=\columnwidth]{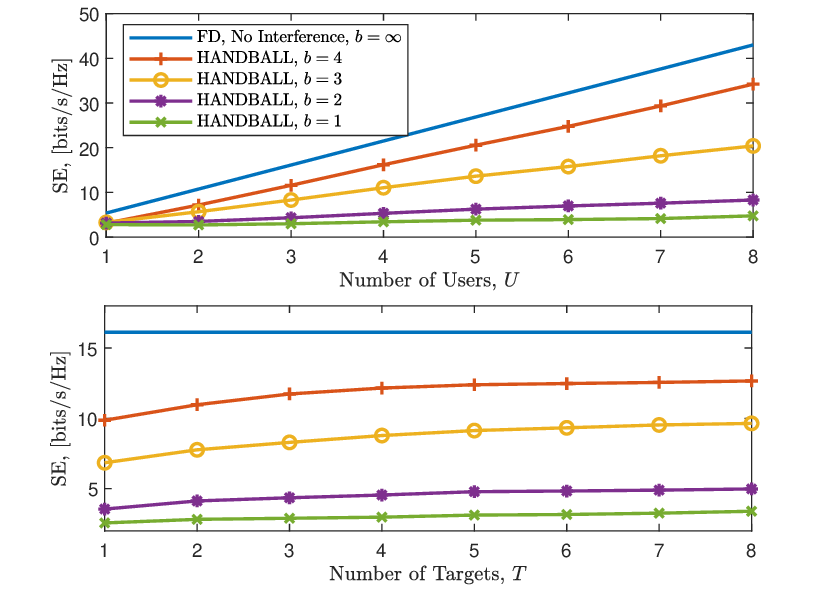} } 
		\caption{SE performance versus $U$ (Top) and $T$ (Bottom) for $\mathrm{SNR} = 10$ dB.
		}
		
		\label{fig_UT}
	\end{figure}

	Fig.~\ref{fig_SNR} shows the SE performance with respect to SNR for $U=3$ and $T=3$ when the communication-sensing trade-off parameter $\eta = 0 .5$. The performance of FD beamformers $\mathbf{f}_{\mathrm{opt},u}$ and $\mathbf{v}_{\mathrm{opt},u}$ with infinite resolution (i.e., $b = \infty$) and no inter-user interference is also shown as a  benchmark.  It can be seen that the performance of our HANDBALL technique improves as the $b \rightarrow \infty$, and it attains a close performance to the FD scheme for approximately $b \geq 4$.  We also present the performance of HANDBALL for communication-only scenario, i.e., $\eta = 1$, which attains the SE of the no-interference case with a slight loss, which is due to the use hybrid (instead of FD) beamformers. We also present the SE performance of the ISAC system with respect to number of users $U$ and number of targets $T$ in Fig.~\ref{fig_UT}. We can see that the SE performance improves with the increase of both $U$ and $T$ because of the increase of number of RF chains as $N_\mathrm{RF} = U + T$. While this increase is more significant for $U$, it is relatively smaller for $T$ because the computation of SE directly relies on the beamformer gain toward the users.

\begin{figure}[t]
		\centering
		{\includegraphics[draft=false,width=\columnwidth]{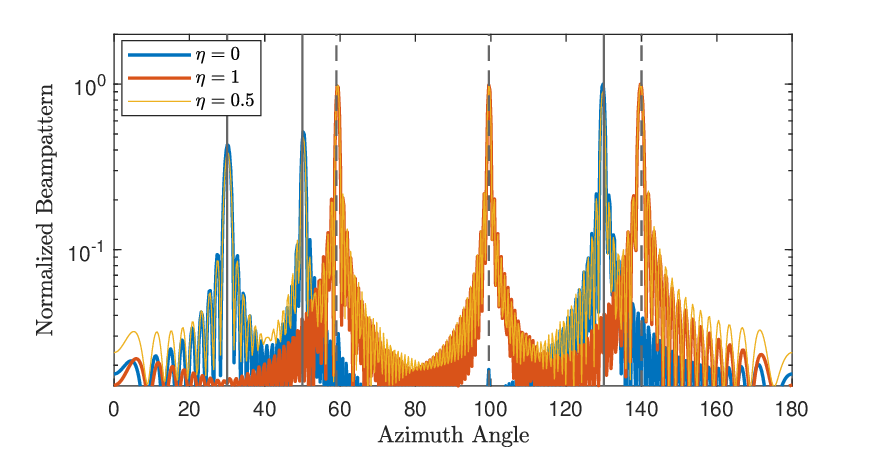} }
		\caption{Normalized beampattern for $\eta \in \{0, 0.5 ,1\}$ when $\mathrm{SNR} = 10$ dB.
		}
		
		\label{fig_bp}
	\end{figure}

	In order to provide further insight, we also present the normalized beampattern of the hybrid beamformers for $b = 1$ in Fig.~\ref{fig_bp}, wherein the LoS targets and users, which are illustrated with straight and dashed lines, with directions $\theta_u \in \{60^\circ,100^\circ,140^\circ\}$ and $\varphi_t \in \{30^\circ,50^\circ,130^\circ\}$, respectively. We can see that the proposed HANDBALL approach simultaneously generates multiple beams toward both users and targets when $\eta = 0.5$. For extreme cases, e.g., $\eta = 0$ ($\eta = 1$), we also observe that the beams are steered toward the targets (users), respectively.

	\section{Conclusions}
	We proposed a hybrid beamforming technique, HANDBALL, for ISAC with low resolution DACs. We have shown that HANDBALL achieves satisfactory performance in terms of both communication as well as sensing in terms of SE and array beampattern, respectively.

	\bibliographystyle{IEEEtran}
	\bibliography{IEEEabrv,references_120}

\end{document}